\documentclass[runningheads]{llncs}
\usepackage[T1]{fontenc}
\usepackage{graphicx}
\usepackage{color}
\usepackage{placeins}
\usepackage{float}
\usepackage{cleveref}
\usepackage{ulem}
\usepackage[table,xcdraw]{xcolor}
\usepackage{array}
\usepackage{textcomp}
\usepackage[font=scriptsize]{subcaption}
\usepackage{adjustbox}
\usepackage{booktabs,array}
\usepackage{tabularx}
\usepackage[font=scriptsize]{caption}

\newcommand{\needscitation}[1]{\textcolor{red}{\textbf{\large Need Citation}}}

\newcommand{\hide}[1]{Redacted until co-authors notification}
\begin{document}

\title{Artifact for A Non-Intrusive Framework for Deferred Integration of Cloud Patterns in Energy-Efficient Data-Sharing Pipelines}
\titlerunning{Deferred Integration of Cloud Design Patterns}
%

\author{
Sepideh Masoudi\inst{1} \and
Mark Edward Michael Daly\inst{1} \and
Jannis Kiesel\inst{1}
}

\authorrunning{S. Masoudi et al.}
%
\institute{
\textsuperscript{1} Information Systems Engineering, Technische Universität Berlin, Berlin, Germany \\
\email{\{smi,jaki\}@ise.tu-berlin.de}\\
\email{m.daly.higham@campus.tu-berlin.de}\\
\url{https://www.tu.berlin/en/ise}
}

\maketitle              
\begin{abstract}
As data mesh architectures grow, organizations increasingly build consumer-specific data-sharing pipelines from modular, cloud-based transformation services. While reusable transformation services can improve cost and energy efficiency, applying traditional cloud design patterns can reduce reusability of services in different pipelines. We present a Kubernetes-based tool that enables non-intrusive, deferred application of design patterns without modifying services code. The tool automates pattern injection and collects energy metrics, supporting energy-aware decisions while preserving reusability of transformation services in various pipeline structures.

\end{abstract}
\section{Introduction}With the emergence of Data Mesh to support data sharing across domains and organizations, data pipelines have become a model for chaining sequential transformations to prepare data for sharing, based on consumer requirements or governance regulations~\cite{machado2022data}. These transformations can be implemented as cloud-based services, benefiting from scalability, flexibility, and efficient practices known as cloud design patterns to improve performance and fault tolerance~\cite{wu2015service}.
Although pipelines are consumer-specific, increasing numbers of consumers allow reuse of transformation services when requirements overlap, e.g., among consumers in the same group~\cite{caise-paper}. Such reuse improves energy efficiency and cost, but requires approaches that support reuse across pipelines with different designs and structures~\cite{caise-paper}.

A key challenge is that different service combinations and pipeline structures demand distinct design patterns. Thus, there is a need to decouple cloud design pattern application from transformation services code. We propose \textit{SnapPattern}\cite{mainPaper}, a tool enabling non-intrusive application of design patterns without modifying service source code. Patterns are integrated automatically via the tool’s interface, allowing Data Mesh administrators to design pipelines from reusable services and dynamically apply patterns. 
In addition, \textit{SnapPattern} monitors the impact of applied patterns on energy consumption and performance, supporting energy-aware decisions in pipeline design. The tool can be integrated into a self-serve data platform to enhance automation, preserve service reusability, and enable informed application of cloud design patterns. In the following, we present the architecture and technical implementation of \textit{SnapPattern}, along with usage instructions and a case study setup description.

\label{introduction}
\section{Tool Structure}
\begin{figure}
\centering
\includegraphics[width=0.80\textwidth]{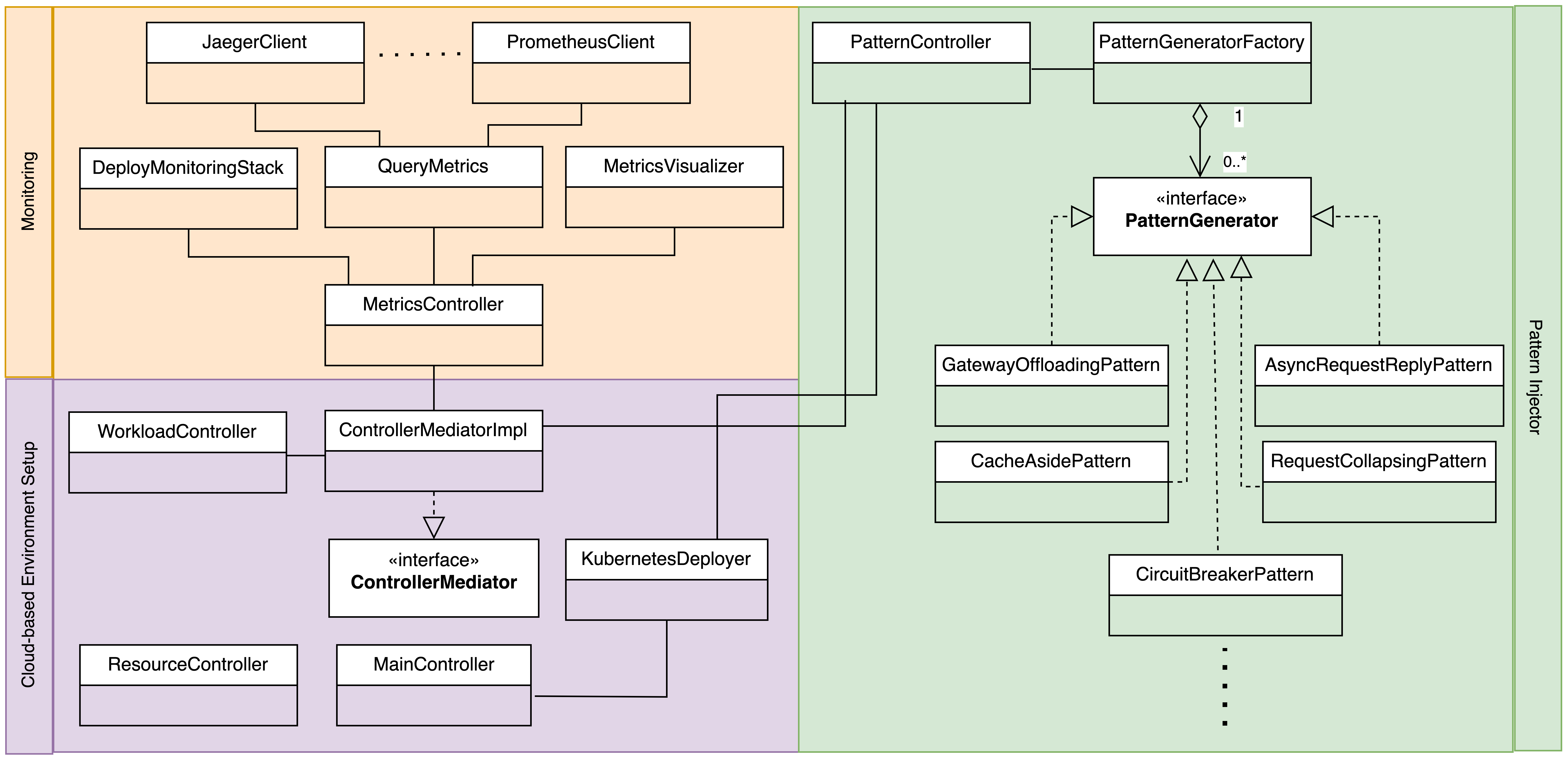}
\caption{Class diagram of SnapPattern.} \label{cDiagram}
\end{figure}

Figure~\ref{cDiagram} presents the SnapPattern class diagram.
The \textit{ResourceController} class provides the list of services across different Kubernetes namespaces to be displayed in the user interface, along with the status of the services.
The \textit{ControllerMediator} is responsible for registering all the controllers that execute the tasks for generating workload, creating the Kubernetes environment and metrics collectiona. 
The \textit{MainController} class serves as the entry point for running the entire program. The user interfaces for each of these components are implemented in the corresponding classes using JavaFX.
At the current stage, \textit{SnapPattern} supports five cloud design patterns: \textit{Circuit Breaker (CB)}, \textit{Cache Aside (CA)}, \textit{Request Collapsing (RC)}, \textit{Gateway Offloading (GO)}, and \textit{Asynchronous Request-Reply (ARR)}~\cite{microsoft_cloud_design_patterns_2022}. 
However, the extension of the design patterns can be easily done by implementing the \textit{SnapPattern} interface and providing the code and resources required for the injection of the new pattern. 
\label{architecture}
\section{Evaluation Methodology and Case Study}To demonstrate SnapPattern’s capability for non-intrusive injection of design patterns in data-sharing pipelines, we implemented four transformation services including filtering, aggregation, anonymization, and formatting, using the Spring Boot framework. As shown in \cref{exp-pipe}, we then defined four pipelines by combining these services in different ways, ensuring reuse across pipelines. These services form the system under test, to which we applied patterns using SnapPattern. The code and deployment files for these transformation services are provided in the paths listed in \cref{tab:repo}.
For the case study, the Circuit Breaker and Asynchronous Request-Reply patterns were applied to all transformation services (Filtering, Aggregation, Anonymization, Formatting), the Gateway Offloading pattern to the Coordinator Service, and the Request Collapsing and Cache Aside patterns to the Data Product service, one pattern for each run of the experiment. After applying the patterns, we generated workloads and collected metrics with SnapPattern, focusing on total energy consumption when running four pipelines under each workload after and before applying pattern. The results were used to generate plots comparing the impact of individual design patterns on the total energy consumption for all pipelines overall.
\subsection{Data origins}
To set up a data-sharing pipeline as a system under test(SUT) for our case study, we used open data from the official Berlin portal, which promotes free data flow, transparency, and fair competition, enabling citizens, businesses, and scientists to create applications and tools\cite{opendata}. The portal provides about 3,104 public datasets from sectors such as traffic, economy, and social benefits, in formats like CSV and JSON. We selected a dataset on banished books\footnote{\url{https://www.berlin.de/verbannte-buecher/suche/index.php/index/all.json?q=}}, as it was available in JSON and had a manageable volume for running the case study on a machine with the specifications described in the next chapter.
\subsection{Case Study Setup Requirements}
For full measurement insights, running the case study evaluation, using the provided data pipeline as the system under test, requires a Linux host. The development environment used for the creation of the artifact comprised an Intel x86\_64 8-core i7-1355U processor with 24~GB of memory. Before starting, the user should install Java~17 and download Apache JMeter, placing JMeter in the main repository path to run workload scripts. For plot generation, Python~3.11 and the dependencies in the requirements file must be installed.
Alternatively, the installation script from the GitHub repository can be used to automatically install Python, the required dependencies, and JMeter.
We applied three workload levels: low, medium, and high, where the number of user requests increases every 30 seconds by 10, 20, and 40 users, respectively.
Maven is used for build automation and dependency management.
\subsection{Case Study Execution Instruction}
\begin{figure}
\centering
\includegraphics[width=0.55\textwidth]{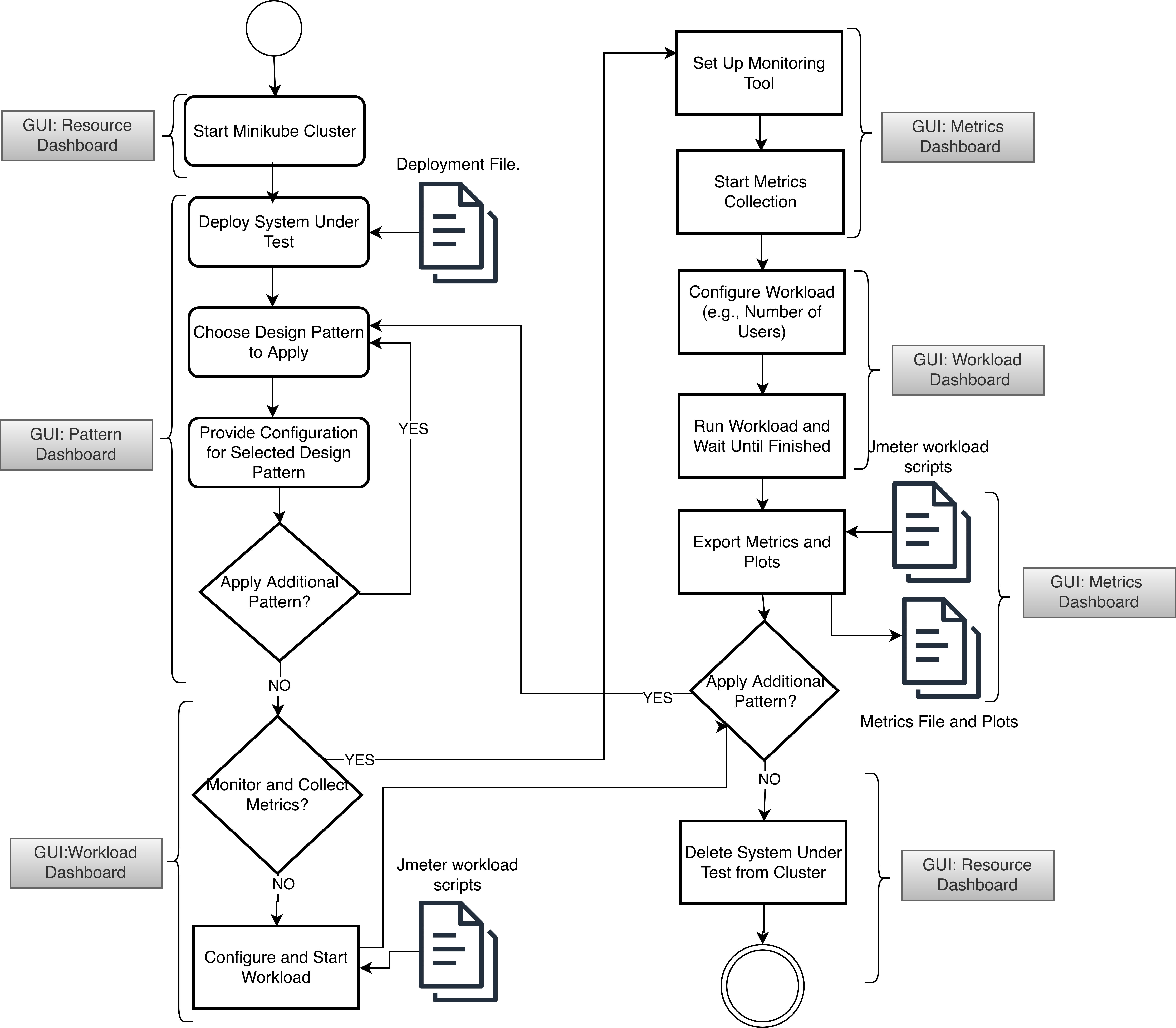}

\caption{Workflow of using SnapPattern.} \label{workflow}
\end{figure}

\begin{figure}[ht]
  \begin{subfigure}{0.48\textwidth}   
    \centering
    \includegraphics[width=\linewidth, trim={-1cm, 2cm, 15cm, 0cm}]{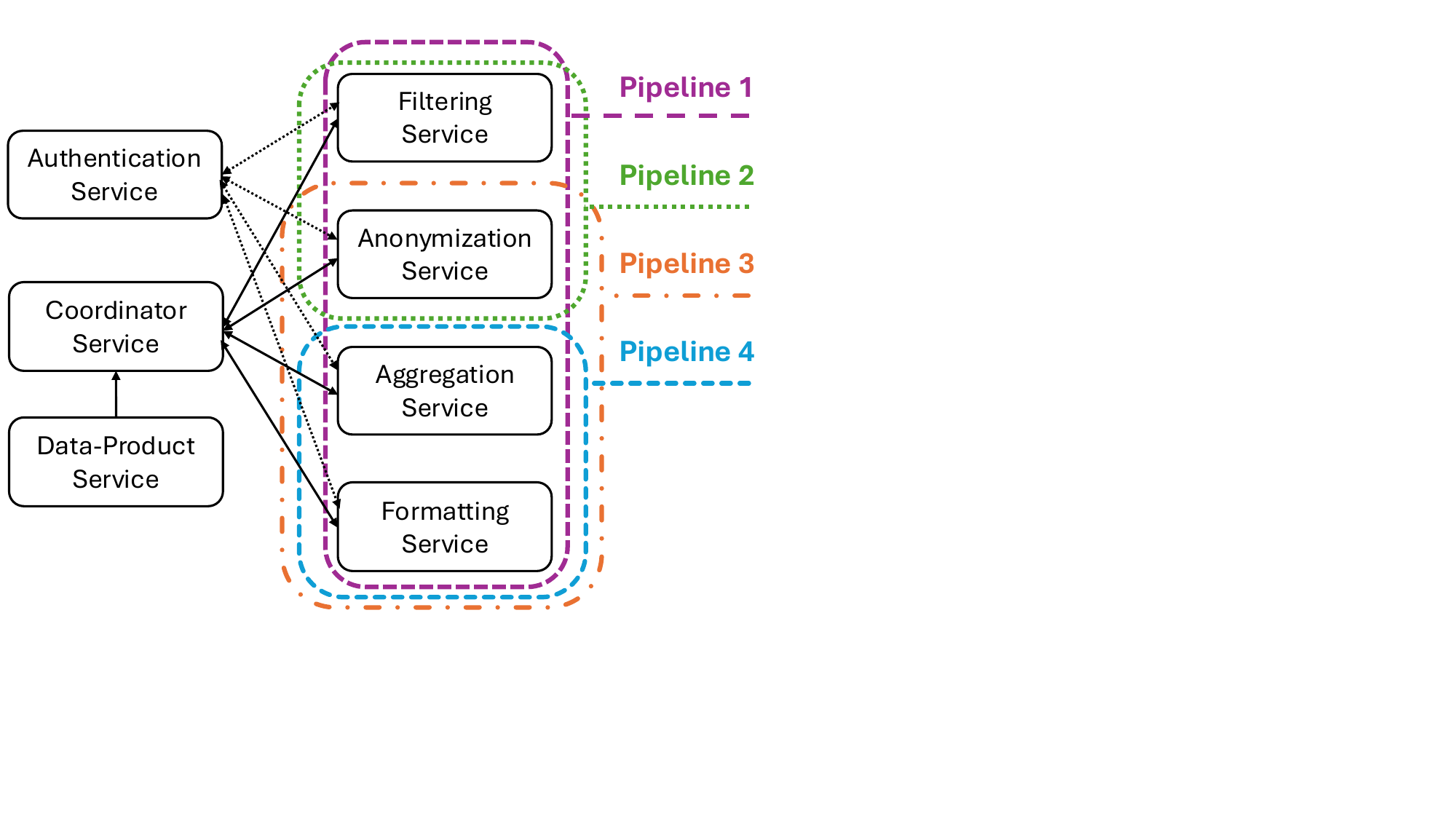}
    \caption{Pipelines constructed using the four transformation stages.}
    \label{exp-pipe}
  \end{subfigure}
  \hfill
  \begin{subfigure}{0.48\textwidth}   
    \centering
   \includegraphics[width=\linewidth]{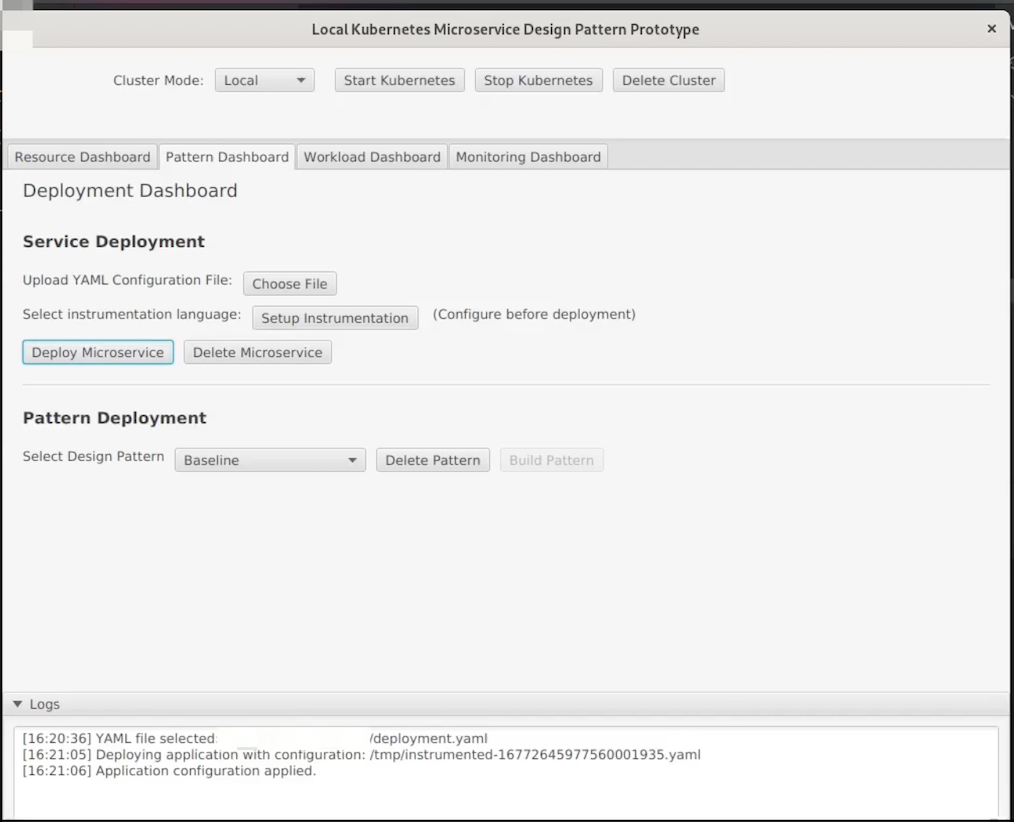}
    \caption{Graphical user interface of the tool.}
    \label{gui-pattern}
  \end{subfigure}

  \caption{Side-by-side view of the experiment pipelines (left) and the tool’s GUI (right).}
  \label{fig:side-by-side}
\end{figure}

To run \textit{SnapPattern}, after preparing the infrastructure as explained, the Java bin path variable must be updated in the WorkloadController.java class, path provided in the README file of the GitHub repository. Next, execute the \texttt{Main.java} file to launch the JavaFX application and display the user interface.
The Graphical User Interface (GUI), shown in \cref{gui-pattern}, consists of four tabs: \textit{Resource}, \textit{Pattern}, \textit{Workload}, and \textit{Monitoring} Dashboards. Each tab is supported by a dedicated controller (see \cref{cDiagram}) to set up the Minikube environment, deploy transformation services and design patterns, generate workloads via JMeter, and collect performance and energy metrics with Prometheus and Kepler, exporting results in CSV and plots respectively.
The case study requires three inputs: the deployment YAML file of the system under test, the JMeter workload script, and the configuration for design pattern injection, all already exist in the repository and ready to use. Following the workflow in \cref{workflow}, all remaining steps, from deploying the cluster and system under test to injecting patterns, deploying the monitoring stack, generating workloads, collecting metrics, and finally deleting patterns, the system, or the cluster can be performed through the corresponding buttons in the GUI. Progress of each step is displayed in the log at the bottom of the GUI. A video tutorial demonstrating the using SnapPattern in the case study is available in the address shown in(\cref{tab:repo}).
The ecxact steps and workflow of using the tool has been demonstrated in \cref{workflow}.
\subsection{Evaluation Configuration Parameters}
In the tables below you can find the configurations that passed to the SnapPattern when injecting each design pattern in our case study.
As mentioned we applied one design pattern at a time to our data-sharing pipelines(SUT).
\begin{table}[H]
\noindent\hspace*{-1.5cm} 
\begin{minipage}[t]{1.05\textwidth} 
\small
\caption{Artifact References}
\begin{tabular}{|l|l|}
\hline
\textbf{Artifact} & \textbf{Address} \\ \hline
SnapPattern Repository & \url{https://github.com/patterninjector/SnapPattern} \\ \hline
Repository README File & \url{https://github.com/patterninjector/SnapPattern/blob/main/Readme.md} \\ \hline
Video Tutorial & \url{https://vimeo.com/1100669519} \\ \hline
Experiment Results &\{SnapPattern Repository\}/Python/results/metrics\_data\_5min\_updated.csv \\  \hline
Data Pipeline Repository (sample SUT) & \url{https://github.com/patterninjector/Data-sharing-pipeline} \\ \hline
Data Source Address & \url{https://www.berlin.de/verbannte-buecher/suche/index.php/index/all.json?q=} \\ \hline
\end{tabular}
\label{tab:repo}
\end{minipage}
\end{table}
\vspace{-1em} 
\begin{table}[H]
\centering
\small
\caption{Circuit Breaker}
\begin{tabular}{|l|l|l|l|l|l|l|l|}
\hline
\textbf{Service} & \textbf{Route} & \textbf{Port} & \textbf{Max Conn} & \textbf{Max Pend} & \textbf{Max Req} & \textbf{Retries} & \textbf{Timeout} \\ \hline
filter-service & / & 8081 & 100 & 20 & 1 & 2 & 1s \\ \hline
formatting-service & / & 8084 & 100 & 20 & 1 & 2 & 1s \\ \hline
aggregation-service & / & 8082 & 100 & 20 & 1 & 2 & 1s \\ \hline
anonymization-service & / & 8083 & 100 & 20 & 1 & 2 & 1s \\ \hline
\end{tabular}
\label{tab:circuitbreaker}
\end{table}
\vspace{-1em} 
\begin{table}[H]
\centering
\begin{minipage}[t]{0.48\textwidth}
\centering
\small
\caption{Async Request Reply}
\begin{tabular}{|l|l|}
\hline
\textbf{Service Name} & \textbf{Endpoint Path} \\ \hline
filter-service & / \\ \hline
formatting-service & / \\ \hline
aggregation-service & / \\ \hline
anonymization-service & / \\ \hline
\end{tabular}
\label{tab:asyncrequest}
\end{minipage}
\hfill
\begin{minipage}[t]{0.48\textwidth}
\centering
\small
\caption{Request Collapsing}
\begin{tabular}{|l|l|}
\hline
\textbf{Field} & \textbf{Value} \\ \hline
Backend Service & data-product-service \\ \hline
Backend Port & 8089 \\ \hline
Endpoint Path & /data-json \\ \hline
Query Parameter & (empty) \\ \hline
ID Field & (empty) \\ \hline
Batch Query & \texttt{\/data\-json} \\ \hline
DB Host & data-product-service.user \\ \hline
DB Port & 8089 \\ \hline
DB Name & (empty) \\ \hline
DB Username & (empty) \\ \hline
DB Password & (empty) \\ \hline
\end{tabular}
\label{tab:requestcollapsing}
\end{minipage}
\end{table}
\vspace{-1em} 
\begin{table}[H]
\centering
\begin{minipage}[t]{0.48\textwidth}
\centering
\small
\caption{Gateway Offloading}
\begin{tabular}{|l|l|}
\hline
\textbf{Field} & \textbf{Value} \\ \hline
Service Name & coordinator \\ \hline
Service Port & 8080 \\ \hline
Service Endpoint & / \\ \hline
\end{tabular}
\label{tab:gatewayoffloading}
\end{minipage}
\hfill
\begin{minipage}[t]{0.48\textwidth}
\centering
\small
\caption{Cache Aside}
\begin{tabular}{|l|l|}
\hline
\textbf{Field} & \textbf{Value} \\ \hline
Backend Service & data-product-service \\ \hline
Backend Port & 8089 \\ \hline
Cached Endpoints & / \\ \hline
Cache TTL (seconds) & 60 \\ \hline
Max Connections & 100 \\ \hline
\end{tabular}
\label{tab:cacheaside}
\end{minipage}
\end{table}
\label{evaluation}
\section{License Information}The source code of our tool and evaluation results are publicly available and released under the MIT License, allowing researchers and practitioners to freely use, modify, and extend it.\label{license}

\begin{credits}
\subsubsection{\ackname} Funded by the European Union (TEADAL, 101070186). Views and opinions expressed are however those of the authors only and do not necessarily reflect those of the European Union. Neither the European Union nor the granting authority can be held responsible for them.
\end{credits}
%
%
%

%
\end{document}